\documentclass{elsart3}

\usepackage{graphicx}
\usepackage{subfigure}
\usepackage{latexsym}

\begin{document}

\newcommand{\greeksym}[1]{{\usefont{U}{psy}{m}{n}#1}}
\newcommand{\umu}{\mbox{\greeksym{m}}}
\newcommand{\udelta}{\mbox{\greeksym{d}}}
\newcommand{\uDelta}{\mbox{\greeksym{D}}}
\newcommand{\uOmega}{\mbox{\greeksym{W}}}
\newcommand{\uPi}{\mbox{\greeksym{P}}}
\newcommand{\ualpha}{\mbox{\greeksym{a}}}

\newcommand{\mrm}{\mathrm}
\newcommand{\Neq}{\mrm{n}_{\mrm{eq}}/\mrm{cm}^{2}}
\newcommand{\fns}{\footnotesize}
\newcommand{\scrs}{\scriptsize}

\sloppy

\begin{frontmatter}

\title{Signal height in silicon pixel detectors irradiated with pions and protons}

\author[psi]{T.\,Rohe\thanksref{corr}},
\author[pr]{J.\,Acosta},
\author[ku]{A.\,Bean}
\author[psi,eth]{S.\,Dambach}
\author[psi]{W.\,Erdmann},
\author[eth]{U.\,Langenegger},
\author[ku]{C.\,Martin},
\author[psi]{B. Meier},
\author[ku]{V.\,Radicci},
\author[ku]{J.\,Sibille},
\author[psi,eth]{P.\,Tr\"ub}
\address[psi]{Paul Scherrer Institut, Villigen, Switzerland}
\address[pr]{University of Puerto Rico, Mayag\"uez}
\address[ku]{University of Kansas, Lawrence KS, USA}
\address[eth]{ETH Z\"urich, Switzerland}
\thanks[corr]{Corresponding author; e-mail: Tilman.Rohe@cern.ch}
\begin{abstract}
Pixel detectors are used in the innermost part of multi purpose experiments
at the Large Hadron Collider (LHC) and are therefore exposed to the highest fluences of ionising
radiation, which in this part of the detectors consists mainly of charged
pions. The radiation hardness of the detectors has thoroughly been tested
up to the fluences expected at the LHC. In case of an LHC upgrade the
fluence will be much higher and it is not yet clear up to which radii the
present pixel technology can be used.
In order to establish such a limit, pixel sensors of the size of one CMS pixel
readout chip (PSI46V2.1) have been bump bonded and irradiated with positive
pions up to $6\times 10^{14}\,\Neq$ at PSI and with protons up to 
$5\times 10^{15}\,\Neq$.
The sensors were taken from production wafers of the CMS barrel pixel
detector. They use n-type DOFZ material with a resistance of about 
$3.7\, \mrm{k}\uOmega\mrm{cm}$ 
and an n-side read out. As the performance of silicon sensors is
limited by trapping, the response to a Sr-90 source was investigated. The
highly energetic beta-particles represent a good approximation to
minimum ionising particles. The bias dependence of the signal for a wide
range of fluences will be presented.

\end{abstract}
\end{frontmatter}

\section{Introduction}

The tracker of the CMS experiment consists of only silicon detectors \cite{ref:cms}. 
The region with a distance to the beam pipe between 22 and 115\,cm is equipped
with  10 layers of single sided silicon strip detectors covering an area
of almost 200\,m$^2$ with about $10^{7}$ readout channels.
The smaller radii are equipped with a pixel detector which was inserted into CMS in 
August 2008. It consists of three barrel layers and two end disks at each side. The 
barrels are 53\,cm long and placed at radii of 4.4\,cm, 7.3\,cm, and 10.2\,cm.
They cover an area of about $0.8\,\mrm{m}^{2}$ with
roughly 800 modules. The end disks are located at a mean distance from the
interaction point of 34.5\,cm and 46.5\,cm. The area of the 96 turbine blade
shaped modules in the disks sums up to about $0.28\,\mrm{m}^{2}$.
The pixel detector contains about $6\times 10^{7}$ readout channels
providing three precision space points up to a pseudo rapidity of 2.1. These unambiguous
space points allow an effective pattern recognition in the dens track environment
close to the LHC interaction point. The precision of the measurement is
used to identify displaced vertices for the tagging of b-jets and $\tau$-leptons.

The two main challenges for the design of the pixel detector are the
high track rate and the high level of radiation. The former concerns the
architecture of the readout electronics while the high radiation level mainly 
affects the charge collection properties of the sensor,
which degrades steadily.

A possible luminosity upgrade of LHC is currently being discussed. With a minor hardware upgrade
a luminosity above $10^{34}\,$cm$^{-2}$s$^{-1}$ might be reached. Later major
investments will aim for a luminosity of $10^{35}\,$cm$^{-2}$s$^{-1}$ \cite{ref:slhc}.
The inner regions of the tracker will have to face an unprecedented track rate
and radiation level. The detectors placed at a radius of 4\,cm
have to withstand the presently unreached particle fluence of $\Phi\approx 10^{16}\,\Neq$
or must be replaced frequently. However, the operation limit of
the present type hybrid pixel system using ``standard'' n-in-n pixel sensors is
not yet seriously tested. The aim of the study presented is to test the charge collection
of the CMS barrel pixel system at fluences exceeding the specified
$6\times 10^{14}\,\Neq$ \cite{ref:tdr}.

\section{Sensor samples}

The sensors for the CMS pixel barrel follow the so called ``n-in-n'' approach. The
collection of electrons is of advantage in a highly radiative environment as they
have a higher mobility than holes and therefore suffer less from trapping. Furthermore, the 
highest electric field after irradiation induced space charge sign inversion is located
close to the collecting n-electrodes. The need of a double sided processing leading to
a significant price increase compared to truly single sided p-in-n sensors is used as a
chance to implement a guard ring scheme keeping all sensor edges on ground potential.
This feature simplifies the design of the detector modules considerably.
For n-side isolation the so called moderated p-spray technique~\cite{ref:mod} has been
chosen and a punch through biasing grid has been implemented.

The sensor samples were taken from wafers of the main production run for the 
CMS pixel barrel which were processed on n-doped DOFZ silicon according to the 
recommendation of the ROSE-collaboration~\cite{ref:rose}. The resistance of
material prior to irradiation was $3.7\,\mrm{k}\uOmega\mrm{cm}$. The 
approximately $285\,\umu$m thick sensors had the size of a single readout chip and 
contain $52\times 80$ pixels with a size of $150\times 100\,\umu\mrm{m}^{2}$
each. In contrast to previous studies \cite{ref:tb-03-04} the standard bump bond 
and flip chip procedure described in~\cite{ref:bump} was applied to the samples.
As this includes processing steps at elevated temperature, this was done before 
irradiation which simplified the whole procedure considerably and resulted in
a very good bump yield. In return it means that the readout chips
were also irradiated. Although the operation of irradiated readout circuits poses 
a major challenge and source of measurement errors, it gives a realistic picture
of the situation in CMS after a few years of running.

The sandwiches of sensor and readout chip were irradiated at the PSI-PiE1-beam line 
with positive pions of momentum 280\,MeV/c to fluences up to 
$6\times 10^{14}\,\Neq$ and with 26\,GeV/c protons at CERN-PS up to 
$5\times 10^{15}\,\Neq$.

All irradiated samples were kept in a commercial freezer at $-18^{\circ}\,$C after 
irradiation. However the pion irradiated ones were accidentally warmed up to room 
temperature for a period of a few weeks (due to an undetected power failure).

\section{Measurement Procedure}

The aim of the study was to determine the amount of a signal caused by
minimum ionising particle (m.i.p.) as a function of sensor
bias and irradiation fluence. For this the response of the samples to a
Sr-90 source was investigated. The endpoint energy of the beta particles 
is about 2.3\,MeV which approximates a m.i.p. well. However
there is also a large number of ``low energy'' particles which are 
stopped in the sensor and cause much larger signals. Those have to be 
filtered during the data analysis.

The samples were mounted on a water cooled 
Peltier element and kept at $-10^{\circ}\,$C. The source was placed inside the 
box about 10\,mm above the sensor. As the compact setup did not allow the
implementation of a scintillator trigger a so called random trigger was used.
In this method the FPGA generating all control signals for the readout chip
stretches an arbitrary cycle of the clock sent to the readout chip by a 
large factor, and, after the latency, sends a trigger to read out the
data from this stretched clock cycle. The stretching factor was adjusted in a
way that about 80\,\% of the triggers showed hit pixels. 

A measurement sequence consists of the following steps:  
\begin{itemize}
\item Cool down the sample while flushing the box with dry nitrogen.
\item The ``pretest'' adjusts basic parameters of the readout chip.
\item The ``full test'' checks the functionality of each pixel.
\item Fine tune the threshold in each pixel to a value of 4000 electrons 
      as uniform as possible (``trim'' the chip). 
\item The pulse height calibration relates for each pixel the pulse height to 
      the DAC values used to inject test pulses. The analogue response is fitted
      to an hyperbolic arc-tangent function \cite{ref:sarah} and the four fit 
      parameters are calculated for each pixel. With procedure an absolute 
      calibration of each pixel is possible.
\end{itemize}
This procedure was identical to what is used to test and calibrate the modules
installed in the CMS experiment. It was perfectly adequate for 
all samples up to a fluence of $1\times 10^{15}\,\Neq$. 

For the samples irradiated to $2.8\times 10^{15}\,\Neq$ the feedback resistor 
of the preamplifier and shaper had to be adjusted manually to compensate for the 
radiation induced change of the transistor's transconductance. The DAC which controls 
this setting is not implemented in the testing software. Then the standard 
calibration procedure was used with the exception that the pixel threshold was 
lowered to about 2000 electrons (instead of 4000). An additional feature of the
readout chip, the leakage current compensation, which might  be useful for such
highly irradiated samples, was not used.

The readout chips of the samples irradiated to $5\times 10^{15}\,\Neq$ 
showed some functionality, however a calibration and quantitative analysis 
of the data was not yet possible and will be the subject of further investigations.
 
After these steps data is taken using the Sr-90 source. The sensor bias was varied
over a wide range. The maximum voltage applied
was 250\,V for the unirradiated samples, 600\,V for the samples irradiated up to 
$1\times 10^{15}\,\Neq$, and 1100\,V for the samples which received a fluence of
$2.8\times 10^{15}\,\Neq$. 
The change of the sensor bias has no effect on the calibration
performed before. The temperature can be kept stable during the bias scan
within $0.2^{\circ}\,$C. The effect of such small temperature variations
has been tested to be negligible.

The data was analysed off line. First all analogue pulse height information
were converted into an absolute charge value, using the parametrisation described
above. After this a pixel mask is generated which excludes faulty pixels. A pixel
was masked if it shows much less (``dead'') or more (``noisy'') hits than its
neighbours, and if the pulse height calibration failed. In addition a manually
generated list of pixels can be excluded. In a second step all clusters of
hit pixels are reconstructed. If a cluster touches a masked pixel or the 
sensor edge, it is excluded from further analysis. Clusters of different
size (one pixel, two pixels, etc.) are processed separately.
To measure the pulse height, the charge of a cluster is summed and
histogrammed. To those histograms a Landau function convoluted with a Gaussian
is fitted. The quoted charge value is the most probable value (MPV) of the Landau. 

Due to the low threshold of only 2000 electrons the highly irradiated samples 
($2.8\times 10^{15}\,\Neq$) showed a higher number of noisy pixels, especially at
the sensor edge where the pixels are larger. However, also some ``good '' pixels
showed a certain number of noise hits which lead to a second peak in the
pulse height spectra. It was well separated from the signal for voltages above 200\,V. 
The origin of the 2 peaks could easily be distinguished:
\begin{itemize}
\item The signal peak moves with higher bias to higher values while the
      noise peak stays at the same position but becomes more prominent (more noise hits 
      at higher bias).
\item The spatial distribution of the signal shows the intensity profile of the
      source, while the noise hits are randomly distributed.
\item The signal peak has a typical Landau shape, while the noise peak is more
      Gaussian.      
\end{itemize}
The quoted signal is again the MPV of a convoluted Landau-Gauss fit.
 
\begin{figure}[t]
\centering
\includegraphics[width=.98\linewidth]{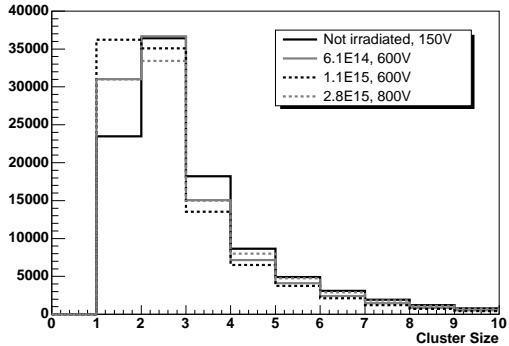}
\caption{Distribution of cluster size for four irradiation fluences.}
\label{fig:clusterSize}
\end{figure}

\begin{figure}[t]
\centering
\includegraphics[width=.98\linewidth]{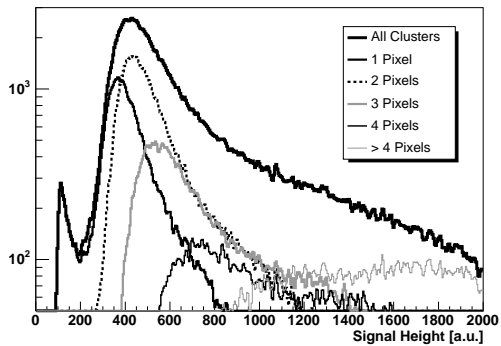}
\caption{Pulse height distribution of an unirradiated sensor in arbitrary units (1 unit
   is about 65 electrons).}
\label{fig:spectrum}
\end{figure}

\section{Results}

Because the radiation of the Sr-90 source contains a large graction of low energy betas
which cause much higher signal than a minimum ionising particle and as the setup was
not equipped with a scintillator which triggered only if a particle penetrated the sample,
the contamination of the low energy particles had to be reduced using the offline analysis.
A particle stopped in the
sensor usually causes part of the ionised electrons to travel in the plane of the 
sensor ionising further electrons in the flight path. This results in large clusters 
of hit pixels. Figure~\ref{fig:clusterSize} shows the distribution of the cluster
size for four irradiation fluences. Naively one would expect a spectrum dominated
by one-hit clusters with a small fraction of clusters of size two to four caused by
particles passing just in-between two pixels or close to a pixel corner. 
However, as visible in Fig.~\ref{fig:clusterSize}, there is a tail of 
events with extremely large clusters, which does not dependent on irradiation or
bias voltage. This supports the hypothesis of secondary particles. Therefore it
is not surprising that the signal is a function of the cluster sizes.
Figure~\ref{fig:spectrum} shows the pulse height distribution of an unirradiated
sensor for different cluster sizes. In particular clusters with more than 4 hit
pixels tend to have very large signals and their distribution can no longer
be described by a Landau function. More surprising is the fact that already
in small clusters with less than four pixels the most probable value of the pulse 
height distribution clearly depends on the cluster size. In order to reduce a contamination
of the data from low energy particles, the pulse height is only extracted
from clusters of size one. 

\begin{figure*}[t]
\includegraphics[width=.98\textwidth]{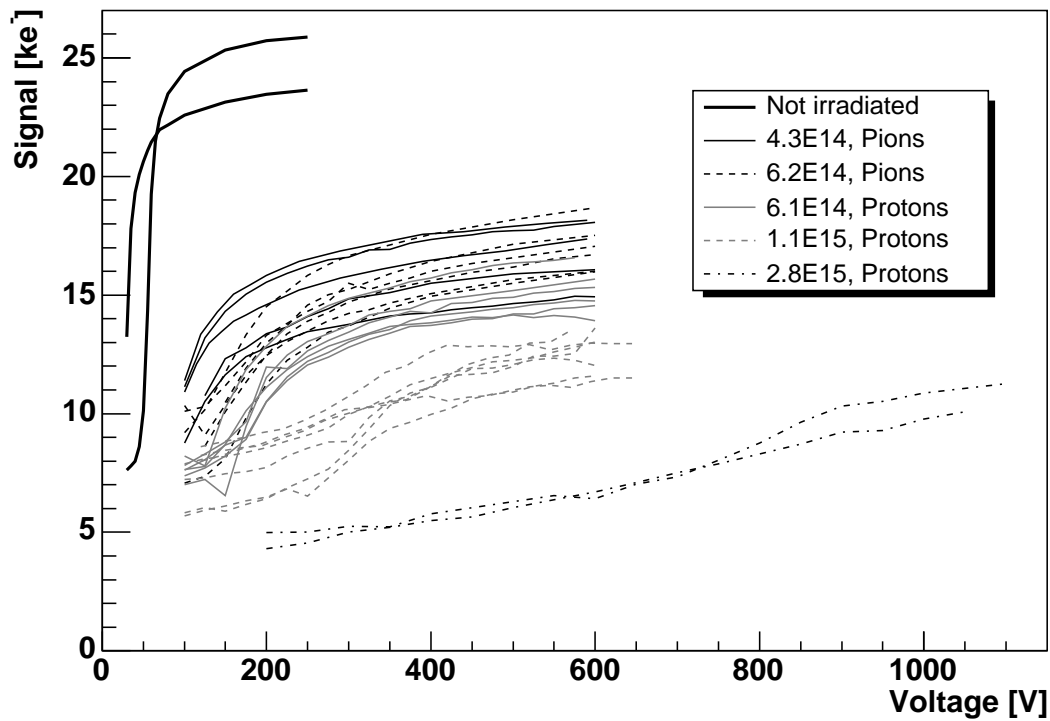}
\caption{Signal from single pixel clusters as a function of the sensor bias. Each line
  represents one sample.}
\label{fig:Q_V}
\end{figure*}

Figure~\ref{fig:Q_V} shows the bias dependence of the signal for all measured samples.
For the unirradiated samples the sudden rise of the signal at the full depletion voltage
of $V_{\mrm{depl}}\approx 55\,$V is nicely visible. The signal then saturates very fast.
The samples irradiated to fluences in the $10^{14}\,\Neq$-range also show a nice saturation
of the signal above roughly 300\,V. The onset of the signal in the ``low'' voltage
range clearly displays the increase of the space charge due to radiation.
There is a strong variation of the saturated signal for samples with the same
irradiation fluence which cannot be explained with differences in the
sensor thickness. The reason for this is probably the imperfection of the pulse height
calibration, which relies on the assumption that the injection mechanism for test pulses
is equal for all readout chips, which is not the case. Variations of the
injection capacitor are larger than $15\,\%$, and also the resistor network in the DAC
shows variations, which are, however, much smaller.
For the samples irradiated to fluences above $10^{15}\,\Neq$, no saturation of the
signal with increasing bias is visible. It is remarkable that even after 
a fluence of $2.8\times 10^{15}\,\Neq$ a charge of more than 10\,000~electrons can
be achieved if it is possible to apply a bias voltage above 800\,V. This nicely 
complements the results for n-in-p strip detectors shown in this 
conference \cite{ref:liv,ref:lj}.

\begin{figure}[t]
\centering
\includegraphics[width=.98\linewidth]{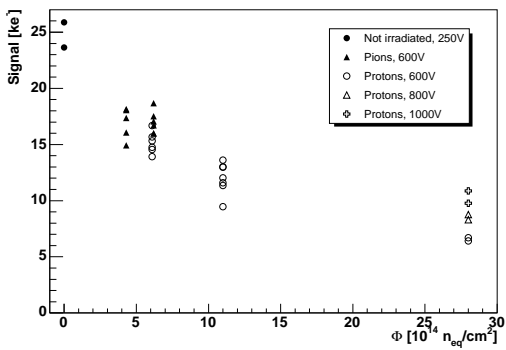}
\caption{Most probable signal as a function of the irradiation fluence. Each point
  represents one sample (apart from the highest fluence where each of the two samples 
  is shown at three bias voltages).}
\label{fig:Q_Phi}
\end{figure}

In order to display the development of the signal height as a function of
the fluence, the charge at 600\,V was extracted for each sample (250\,V for 
the unirradiated ones) and plotted in Fig.~\ref{fig:Q_Phi}. In addition the 
values for 800\,V and 1000\,V are plotted for the highest fluence.
Apart from the large fluctuations, which are due to the calibration
of the readout electronics, the reduction of the charge with fluence is
nicely visible. Further it becomes obvious that it pays to go to 
very high bias voltages if the fluences exceeds $10^{15}\,\Neq$.

\section{Conclusion}

In order to estimate the survivability of the present CMS barrel pixel detector
in a harsh radiation environment, single chip detectors (sensors bump bonded to a
readout chip) have been irradiated to fluences up to 
$5\times 10^{15}\,\Neq$ and tested with a Sr-90 source. The samples
that received fluences up to about $10^{15}\,\Neq$ could be used without
any modification of the chip calibration procedure and obtained a signal charge
of above 10\,000 electrons at a bias voltage of 600\,V. From this point of 
view their performance is perfectly adequate for the CMS experiment, even
at fluences twice as high as the $6\times 10^{14}\,\Neq$ specified in the Technical design 
report~\cite{ref:tdr}. The samples irradiated to $2.8\times 10^{15}\,\Neq$ could be operated
with slightly adjusted chip settings and also showed a signal of about 10\,000 electrons,
however at a bias voltage of 1000\,V. This indicates the suitability of such devices
for a use at an upgraded LHC. The samples which received $5\times 10^{15}\,\Neq$ could
not yet be operated. Their examination is subject of further studies.

\section*{Acknowledgement}

The pion irradiation at PSI would not have been possible without 
the beam line support by Dieter Renker and Konrad Deiters, PSI, the 
logistics provided by Maurice Glaser, CERN, and the great effort
of Christopher Betancourt and Mark Gerling, UC Santa Cruz (both were supported by 
a financial contribution of RD50 and PSI).

The proton irradiation was carried out at the CERN irradiation facility.
The authors would like to thank Maurice Glaser and the CERN team for the outstanding
service.

The work of J.\,Acosta, A.\,Bean, C.\,Martin, V.\,Radicci and J.\,Sibille
is supported by the PIRE grant OISE-0730173 of the US-NSF.

The work of S.\,Dambach, U.\,Langenegger, and P.\,Tr\"ub is
supported by the Swiss National Science Foundation (SNF).

The sensors  were produced by CiS GmbH in Erfurt, Germany.
\bibliographystyle{elsart-num}
\bibliography{bib_rohe}

\end{document}